# Dynamical characteristics of a Hodgkin-Huxley neuron

Huijie Yang[1]*, Fangcui Zhao[1], Yizhong Zhuo[2], Xizhen Wu[2], Zhuxia Li[2]
1 Physics Department, Hebei University of Technology, Tianjin300130, China
2 China Institute of Atomic Energy, P.O. Box 275(18), Beijing 102413, China

**Abstract**

By means of the concepts of factorial moment, return map and NM estimator, we analyze some responses of a HH neuron to various types of spike-train inputs. The corresponding fractal dimensions and values of NM estimators can describe the correlation characteristics quantitatively. It is found that all the intervals of successive output spikes may obey a power-law for some conditions. And in time scale the output series for a HH model for each type of inputs is correlated with typical correlation length and correlation strength. These power-law and correlation characteristics may be useful in the coding and decoding process of HH neurons. These concepts may distinguish different inputs quantitatively.

---

* Corresponding author, E-mail: huijieyangn@eyou.com

# I. Introduction

Measurement records for a complex dynamical process form a long time series. And analysis of this time series will reveal the essential dynamical mechanics for the process, which in turn can provide us physical scenarios and criteria to construct dynamical models. Many natural sequences have attracted special attentions recently. Typical examples include DNA sequences, weather and climate records, heartbeat and gait series, price evolutions and information streams in Internet, etc. [1-15]. A common feature for all these sequences is that there exist a long-range correlation for the elements, that is to say, the correlation function obeys a power-law and there is not a finite correlation length. In this paper we try to analyze some response time series of a HH neuron to various types of spike-train inputs, which maybe helpful for us to understand the dynamical mechanism for neuron's coding and decoding process.

Neurons in our brain are known to be responsible for encoding the characteristics of stimuli into a form for further processing by other neurons. Anatomical, physiological and theoretical studies on neurons have been made extensively for several decades, but the coding mechanism has not been clarified at the moment [16]. It is commonly believed that the firing rate reflects the strength of the inputs that trigger the action potentials of neurons. Indeed, the firing activities of motor and sensory neurons vary in response to the applied stimuli. It is not known, however, whether the information is carried through the mean firing rate (rate encoding) or through the details of sequences of the temporarily encoded inter-spike interval (ISI) (temporal encoding), which is currently controversial [17-19]. In the last few years, experimental evidences have accumulated, indicating that many biological systems use the temporal coding. Human visual systems, for example, have shown to classify patterns within 150msec in spite of the fact that at least ten synaptic stages are involved from retina to the temporal brain [20]. The similar speed of visual proceeding has been reported for macaque monkeys [21]. Because the firing frequency of neurons involved is less than 100 Hz, each neuron can contribute at most one or two spikes to such computations; there is not sufficient time to sample firing rates [22].

Although much of debates on the nature of the neural code has focused on the rate versus temporal codes, there is the other important issue to be considered: information is encoded in the activity of single (or very few) neurons or that of a large number of neurons (population or ensemble code). The population rate code model assumes that information is coded in the relative firing rates of ensemble neurons, and has been adopted in the most theoretical analysis. On the contrary, in the population temporal code model, it is assumed that relative timings between spikes in ensemble neurons may be used as encoding mechanism for perceptual processing [18,23-24]. A number of experimental data supporting this code have been reported in recent years [25-28]. It has been demonstrated that temporally coordinated spikes can systematically signal sensory object feature, even in the absence of changes in firing rate of the spikes [26,29].

Many studies on the encoding of the spike trains by neurons have been made by using the IF model (integrate and fire model) [30-34], FN model [35-40] and HH model [22,29, 41-45]. It is pointed that HH model is much more realistic comparing with other models. Citing a typical feature, the oscillation of the HH neuron may become chaotic when the sinusoidal $I_s$ is applied with proper choices of magnitude and frequency [46-47], which are experimentally observed in squid giant axons and Onchidium neurons [48-50]. Various concepts are proposed to investigate the characteristics of responses of neurons to various types of spike train inputs, such as the return map, the peak-to-width ratio of output signals, SNR, the cross-correlation between input and output signals, the mutual information, etc. [22,33,38,39,40,42,43,44]. Analytical tools, such as the Fourier transformation (FT), the short-time Fourier transformation (STFT), the detrended fluctuation approach (DFA) and the wavelet transformation (WT), are employed to study the dynamical processes of neurons. It is believed that the chaotic behavior and the stochastic resonance (SR) may be essential for neurons' coding and decoding mechanism [51-62]. Describing the dynamical characteristics of HH neurons completely and quantitatively is still an essential role for us to understand the dynamical processes of neurons at present time.

As a signal transmission system, a signal will transmit from one layer to the next layer, and so on. The dynamical characteristics acting as signal carriers should possess some special features, such as stability and sensitivity to the change of stimuli. That is to say, these dynamical characteristics should not attenuation distinctly during the recursive self-similar encoding procedures from one to another layer. What is more, these dynamical characteristics should be sensitive to the initial stimulus. Therefore finding more dynamical indices with good characteristics is the first step to find signal carriers in HH neurons.

In reference [22], the author has discussed the return maps, the histograms, the means and root mean squares of the responses of a HH neuron to various types of inputs. The surrogate data method is also used to investigate the nature of correlations in output sequences. It is pointed that analysis of the time correlation of the ISI data, for example by plotting return maps can make the distinction of the responses to various inputs. Part of the simulations is generated here. And for the first time, we suggest in this paper the concept of factorial moment and the concept of NM estimator to describe quantitatively the correlation characteristics of the responses of a HH neuron to various types of spike-train inputs. Just as shown in the next paragraph, factorial moments and NM estimator can provide us a quantitative description for correlations in a temporal and/or spatial series. Together with other indices we expect to obtain a much more legible landscape of a HH neuron's dynamical process.

Our paper is organized as follows. In section *II*, the Hodgkin-Huxley (HH) neuron model is mentioned briefly. The concept of factorial moment and the NM estimator are introduced in a simple way. In section *III*, responses of a HH neuron to time-independent ISIs and time-dependent ISIs modulated by sinusoidal signal are investigated by means of return maps, the factorial moments, and the Nm estimators, respectively. What is more, the effect of Gaussian noise is also calculated. The final section *IV* is devoted to conclusion and discussion.

## II. Calculation Methods
**A. Hodgkin-Huxley neuron model** [22,63]

Consider a simple system consisting a HH neuron and a synapse. The synapse is described by an alpha function. In HH model the nonlinear coupling among the four variables, e.g., $V$ for the membrane potential and $m, h, n$ for the gating variables of $Na$ and $K$ channels, can be described by a set of differential equations as follows,

$$C\frac{dV}{dt} = -g_{Na}m^3h(V-V_{Na}) - g_Kn^4(V-V_K) - g_L(V-V_L) + I_p + I_s, \quad (1)$$

$$\frac{dm}{dt} = -(a_m + b_m)m + a_m, \quad (2)$$

$$\frac{dh}{dt} = -(a_h + b_h)h + a_h, \quad (3)$$

$$\frac{dn}{dt} = -(a_n + b_n)n + a_n, \quad (4)$$

where

$$a_m = \frac{0.1(V+40)}{1-e^{-(V+40)/10}}, \quad (5)$$

$$b_m = 4e^{-(V+65)/18}, \quad (6)$$

$$a_h = 0.07e^{-(V+65)/20}, \quad (7)$$

$$b_h = \frac{1}{1+e^{-(V+35)/10}}, \quad (8)$$

$$a_n = \frac{0.01(V+55)}{1-e^{-(V+55)/10}}, \quad (9)$$

$$b_n = 0.125e^{-(V+65)/80}. \quad (10)$$

And $I_s$ expresses the static dc current. $I_p$ denotes the pulse current induced by the spike-train inputs.

Consider the delta-function-type spike-train input expressed by,

$$U_i(t) = V_a \sum_n \delta(t - t_{in}). \quad (11)$$

The firing time $t_{in}$ for arbitrary n is defined by,

$$t_{in+1} = t_{in} + T_{in}(t_{in}), \quad (12)$$

$$t_{i1} = 0, \quad (13)$$

Assuming this spike-train input is injected through the synapse, the current $I_p$ can then be written as,

$$I_p(t) = g_{syn} \sum_n \alpha(t - t_{in})(V_a - V_{syn}), \quad (14)$$

$$\alpha(t) = (t/\tau)e^{-t/\tau}\Theta(t), \quad (15)$$

$$\Theta(t) = \begin{cases} 1, (t > 0) \\ 0, (t < 0). \end{cases} \quad (16)$$

Here the reversal potentials of $Na, K$ channels and leakage are $V_{Na} = 50mV$, $V_K = -77mV$, and $V_L = -54.5mV$; the maximum values of corresponding conductivities are $g_{Na} = 120mS/cm^2$, $g_K = 36mS/cm^2$, and $g_L = 0.3mS/cm^2$; the capacity of the membrane is $C = 1\mu F/cm^2$; $V_a = 30mV$, the reversal potential of synapse $V_{syn} = -50mV$, and the time constant relevant to the synapse conduction $\tau = 2m\sec$. The conductivity of synapse $g_{syn}$ is treated as a parameter.

When the membrane potential $V$ oscillates, it yields the spike-train output, which may be expressed by,

$$U_o(t) = V_a \sum_k \delta(t - t_{ok}), \tag{17}$$

and the output ISI is given by,

$$T_{ok} = t_{ok+1} - t_{ok} \tag{18}$$

Actually, all the neurons work in noisy environment, and the noises can be considered by adding a random current term $I_n$ in the right side of equation (1).

The differential equations for HH model are solved by the fourth Runge-Kutta method for *4sec* with the integration time step *0.01msec*. A temporal sequence with several hundreds elements is long enough for us to get a reliable factorial moment result.

**B. Factorial moment (FM)** [10-14,64]

More than ten years have witnessed a remarkably intense experimental and theoretical activity in search of scale invariance and fractal in multihardron production processes, for short also called "intermittency" [64]. Generally, intermittency can be described with the concept of probability moment (PM). Dividing a region of phase space $\Delta$ into $M$ bins, the volume of one bin is then $\delta = 1/M$. And the definition of $q$-order PM can be written as,

$$C_q(\delta) = \sum_{m=1}^{M} p_m^q, \tag{19}$$

Where $p_m$ is the probability for a particle occurring in the $m$'th bin, which satisfies a constrained condition, $\sum_{m=1}^{M} p_m = 1$. For a self-similar structure, PM will obey a power law as,

$$\lim_{\delta \to 0} C_q(\delta) \propto \delta^{(q-1)D_q}, \tag{20}$$

And $D_q$ is called $q$-order fractal dimension or Renyi dimension. Simple discussions show that $D_0$, $D_1$ and $\{D_q|q \geq 2\}$ reflect the geometry, information entropy and particle correlation dimensions, respectively.

It is well known that intermittency is related with strong dynamical fluctuations. But the measurements for multihardron production obtain the distribution of particle numbers directly instead of the probability distribution. And the finite number of cases will induce statistical fluctuations. To describe the strong dynamical fluctuations and dismiss the statistical fluctuations effectively, factorial moment (FM) is suggested to investigate intermittency [65-66]. The generally used form for FM can be written as,

$$F_q = M^{q-1} \sum_{m=1}^{M} \frac{<n_m(n_m-1)...(n_m-q+1)>}{<n>^q}, \quad (21)$$

Where M is the number of the bins the considered interval being divided into, $n_m$ the number of particles occurring in the $m$'th bin, and $n$ the total number of particles in all the bins. A measure quantity can then be introduced to indicate the dynamical fluctuations,

$$\phi_q = \lim_{\delta \to 0} \frac{\ln F_q}{\ln(1/\delta)}. \quad (22)$$

Here we present a simple argument for the ability of FM to dismiss statistical fluctuations.

The statistical fluctuations will obey Bernoulli and Poisson distributions for a system containing uncertain and certain number of total particles, respectively. For a system containing uncertain total particles, the distribution of particles in the bins can be expressed as,

$$Q(n_1, n_2, ...n_M | p_1, p_2, ...p_M) = \frac{n!}{n_1! n_2! ... n_M!} p_1^{n_1} p_2^{n_2} ... p_M^{n_M}. \quad (23)$$

And $(p_1, p_2, ...p_M)$ are the probabilities for a particle occurring in the $1, 2, ..., M$ bins, respectively. Hence,

$$\langle n_m(n_m-1)...(n_m-q+1) \rangle$$
$$= \int dp_1 dp_2 ... dp_M P(p_1, p_2, ...p_M) \times \sum_{n_1} ... \sum_{n_M} Q(n_1, n_2, ..., n_M | p_1, p_2, ...p_M)$$
$$\times n_m(n_m-1)...(n_m-q+1)$$

$$= n(n-1)...(n-q+1) \times \int dp_1 dp_2 ... dp_M P(p_1, p_2, ..., p_M) p_m^q$$

$$= n(n-1)...(n-q+1)\langle p_m^q \rangle. \tag{24}$$

That is to say,

$$F_q(M) = C_q(M) \propto M^{\phi_q}, |M \to \infty. \tag{25}$$

Therefore FM can describe the strong dynamical fluctuations and can dismiss the statistical fluctuations due to finite number of cases effectively.

The successive output ISIs can form a temporal series, which can be written as $\{T_{ok}|k=1,2,...K\}$. For each type of input spike-train the values of output ISIs vary in a special interval. Firstly, divide this interval into $M$ bins and reckon the number of output ISIs occurring in each bin, respectively. And then calculate the corresponding $q$-order FM, e.g., $F_q(M)$. Iterations of this procedure with different $M$ values will provide us the relation between FM and $M$. If there exist a self-similar structure in the distribution of the values of the output ISIs, the curve of $LnF_q$ versus $LnM$ will be a straight line. Denoting the slope of this straight line with $\phi_q$, the corresponding $q$-order fractal dimension is $D_q = \frac{\phi_q}{q-1}$. However, this kind of fractal structure is independent with the temporal order of the output ISIs. That is to say, shuffling the temporal series, $\{T_{ok}|k=1,2,...K\}$, we can obtain same results for $LnF_q$ versus $LnM$. Hence, FM results obtained with this procedure are called time-independent FM in this paper.

To obtain FM results reflecting the characteristics of the temporal order of the output ISIs, we study an integrated temporal series instead of the initial one as $\{S_{ok}|k=1,K\}$, where $S_{ok} = \frac{1}{k}\sum_{l=1}^{k} T_{ol}$. Obviously, shuffling the initial temporal series, we will obtain different results for $LnF_q$ versus $LnM$. Accordingly, FM results obtained with this procedure are called time-dependent FM in this paper [67].

With the increase of strength of long-range correlation, $q$-order dimensions will become bigger and bigger. Therefore, $q$-order dimension can be regarded as a quantitative measurement for long-range correlations.

### C. NM estimator [68-70]

NM estimator is initially designed to distinguish chaotic behaviors from statistical fluctuations and measurement errors. And it is also successfully used in analysis of DNA chains to find coding regions. Construct a process as illustrated below,

(a) $d$ successive elements are regarded as a case containing d particles. The state of the case can be described with a $d$-dimensional vector as $(x_1, x_2, x_3...x_d)$, where $x_i$ is the state value for the $i'th$ element.

(b) For a temporal series containing $M$ elements, the total possible $M-d+1$ successive cases form a process. This process covers the entire temporal series we are interested in, which can be expressed with the series in $d$-dimensional delay-register vectors:

$$(x_1, x_2, x_3...x_d)$$

$$(x_2, x_3, x_4...x_{d+1})$$

$$\downarrow$$

$$(x_{M-d+1}, x_{M-d+2}, x_{M-d+3}...x_M) \quad (26)$$

Choosing an arbitrary case as reference, NM estimator can be defined as,

$$C_{NM}^0(d) = \frac{1}{\sqrt{(M-d)d}} \cdot \sqrt{\sum_{k=1}^{M-d}\sum_{l=1}^{d}(x_{l+k}-x_l)^2} \quad (27)$$

If the elements are distributed randomly, the NM estimator $C_{NM}^0(d)$ should be large and keep constant for different values of size $d$. However, if there exist a deterministic structure embedded in the considered temporal series, $C_{NM}^0(d)$ will decrease with the increasing of the value of size $d$ and tend to an asymptote lower than that for random distributed temporal series. Therefore, NM estimator can describe a chaotic or correlation behavior quantitatively. To compare NM estimators for the temporal series constructed with different output ISIs a transformation is defined as,

$$C_{NM}(d) = 10C_{NM}^0(d)/\max[C_{NM}^0(d)]. \quad (28)$$

When the size $d$ is smaller than the correlation length $\lambda$ embedded in the temporal series, $C_{NM}(d)$ will decrease rapidly with the increase of $d$. And $C_{NM}(d)$ will tend to an asymptote when $d$ is bigger than the correlation length $\lambda$. Therefore $C_{NM}(d)$ can tell us not only the strength of correlations, but also the correlation length $\lambda$.

## III. Calculation Results

### A. Time-independent input ISI

The responses of a HH neuron to several spike-train inputs with $I_s = 25\mu A/cm^2$ and $T_i = 10, 15, 26, 30 m\sec$ are calculated, respectively. When $T_i$ is small ($\sim 10 m\sec$), the output ISIs are forced to be the same, $T_{om} = T_i$. With the increase of $T_i$, the output ISIs distribute in a special interval for each $T_i$. Though the means of $T_{ok}$ are nearly equal to 10.75$msec$ for all values of $T_i$, the series of $\{T_{ok} | k = 1, K\}$ have different characteristics completely, which can be seen easily in the FM results and NM estimator values. In Fig.(1), a typical result is depicted for $T_i = 30 m\sec$.

From the slopes of FM, e.g., $LnF_q$ versus $LnM$, the corresponding fractal dimensions can be calculated with the relation $D_q = \dfrac{\phi_q}{q-1}$, as shown in table (1). The asymptote values of NM estimators for different inputs are also presented. We can find that the slopes of time-dependent FM are all equal to zero. Hence there should not exist long-range correlations for the output series in time scale. NM estimator can also tell us that there are strong short-range correlations embedded in the output series in time scale. The correlation length can be estimated to be 20 spikes for the time-independent input ISIs considered here, though the correlation strength is dependent strongly on $T_i$. There is not self-similar structure, but there exist deterministic structure in the condition $T_i = 15 m\sec$. But the values of the intervals of successive spikes of outputs should obey a long-range correlation law for some special $T_i$ values large enough ($\sim 26, 30 m\sec, etc.$). And the $q$-order fractal dimension values vary with different values of $q$, e.g., multi-fractal structures exist in these output ISIs.

Results for $I_s = 0\mu A/cm^2$, e.g., the characteristics of a silent neuron are obtained in our calculations. Just as described in reference [22], exact periodic oscillations occur in the output ISIs for various types of spike-train inputs.

### B. Time-dependent Input ISI: sinusoidal modulation

Consider a spike-train input whose ISI is modulated by the sinusoidal signal given by,

$$T_{in} = d_0 + d_1 \sin(2\pi t/T_p), \tag{29}$$

where $T_p, d_0$ and $d_1$ are the period and the coefficients used to adjust the mean and the root-mean-square of the output ISIs, respectively.

Results for $(I_s, d_0, d_1) = (0,20,10), (25,20,10)$ and $(25,10,5)$ are calculated. The corresponding q-order fractal dimensions and NM estimator values are also shown in Table (1). We can find that the fractal dimensions are different for the three conditions. And the NM estimator values are different completely. There is not deterministic structure, but there exist self-similar structures in the condition $(I_s, d_0, d_1) = (0,20,10)$. NM estimators may be much more sensitive to the parameters $(I_s, d_0, d_1)$ rather than the fractal dimensions.

**Table (1) Characteristics for output ISIs**

| $I_s (\mu A/cm^2)$ | $T_i (m\sec)$ | Slope of time-independent FM | | | Slope of time-dependent FM | | | NM estimator | |
|---|---|---|---|---|---|---|---|---|---|
| | | q=2 | q=3 | q=4 | q=2 | q=3 | q=4 | Strength | Length |
| 25 | 10 | 1.00 | 1.00 | 1.00 | 0.00 | 0.00 | 0.00 | 0.83 | 20 |
| 25 | 15 | 0.00 | 0.00 | 0.00 | 0.00 | 0.00 | 0.00 | 2.60 | 20 |
| 25 | 26 | 0.38 | 0.47 | 0.52 | 0.00 | 0.00 | 0.00 | 1.35 | 20 |
| 25 | 30 | 0.54 | 0.64 | 0.66 | 0.00 | 0.00 | 0.00 | 1.40 | 20 |
| 25+Noises | 30 | 0.40 | 0.46 | 0.50 | 0.00 | 0.00 | 0.00 | 1.38 | 20 |
| 25 | $20+10\sin(2\pi t/100)$ | 0.20 | 0.27 | 0.31 | 0.00 | 0.00 | 0.00 | 1.46 | 20 |
| 25 | $10+5\sin(2\pi t/100)$ | 0.27 | 0.31 | 0.32 | 0.00 | 0.00 | 0.00 | 6.95 | 8 |
| 0 | $20+10\sin(2\pi t/100)$ | 0.13 | 0.24 | 0.33 | 0.00 | 0.00 | 0.00 | 8.50 | 1 |

**C. Effect of Gaussian noise** [29,71-73]

It is interesting to analyze the response of a HH neuron to spike-train inputs in noisy environment. There are four resources of noises: (i) cells in sensory neurons are exposed to noises arising from outer world; (ii) ion channels of the membrane of neurons are stochastic;(iii) the synaptic transmission yields noises from random fluctuations of the synaptic vesicle release rate; and (iv) synaptic inputs include leaked currents from neighboring neurons. In literatures the items (i)-(iii) and (iv) are simulated with Gaussian noises and Poisson/Gamma spike-train noises, respectively.

Here we consider the terms (i)-(iii) with Gaussian noise given by [29],

$$\langle \overline{I_n(t)} \rangle = 0, \qquad (30)$$

$$\langle \overline{I_n(t)I_l(t')} \rangle = 2D\delta(n-l)\delta(t-t'), \qquad (31)$$

where $\langle X \rangle$ and $\overline{X}$ are spatial and temporal averages, respectively, $D$ the intensity of white noises. $D=1$ in present simulations.

Results for $(I_s, T_i, D) = (25, 30, 1)$ are presented in Table (1). Comparing with the condition $(I_s, T_i, D) = (25, 30, 0)$ we can find that noises may induce dynamical patterns different completely with that without noises. The fractal dimensions may be much more sensitive to the external noises than the NM estimators.

In Fig.(2) we can find an interesting fact that the FM results for the condition with noises obey a power-law in the scale range $M = 2 - 64$, and are coincide with that for the condition without noises exactly. In the scale range $M = 64 - 128$, the FM results do not increase at all, while for $M > 128$ a power-law is obeyed. That is to say, contributions from noises can be neglected for $M < 64$, while they are essential for $M > 64$. And two mutation points exist. As a result, a platform appears in each curve $LnF_q$ versus $LnM$ for the conditions noises are present and crossover phenomenon appears, which is similar with that in DFA (detrended fluctuation method) in a certain degree [74-77]. Therefore, noises can provide a new mechanism in a HH neuron's dynamical process. Environment noises make the dynamical patterns of a HH model much more rich.

## IV. Conclusion and Discussion

By means of several concepts in analyzing temporal series we investigate quantitatively the responses of a HH model to various types of spike-train inputs. Accompanying with the return map suggested in reference [22], the *q*-order fractal dimensions and the NM estimator could describe correlations in dynamical process of a HH neuron quantitatively and completely.

We can find rich dynamical patterns for the conditions $I_s \neq 0 \mu A/cm^2$, though the means and the root-mean-squares of output ISIs are irrespective of the values of input ISIs.

We can also find that noises may play an essential role in the dynamical process of a HH neuron's coding and decoding mechanism. Effectiveness of noises on the dynamical process of a HH neuron maybe much more important in considering the encoding mechanisms.

There are not long-range correlations in time scale for output temporal series, e.g., there exist a finite correlation length for each output series in this scale. There is a self-similar structure in the distribution of the values for each output series. Hence, the dynamical characteristics of the output ISIs (the dynamical patterns, such as long-range correlations, deterministic structures, etc.) rather than the simple mean ratio may take an essential role in this dynamical process. The transmission characteristics of these dynamical indices for HH neuron ensemble will help us to find the signal carriers in neuron networks, which will be investigated further.

## ACKNOWLEGEMENT

This work was partly supported by the National Science Foundation of China under Grant No.19975073, Grant No. 10175093 and Grant No. 10175093. Yizhong Zhuo thanks also the National Science Foundation of China under Grant No.10075007, Grant No.10175095 for support. Fangcui Zhao thanks also the Youth Science Foundation of Hebei university of Technology under Grant No.Q200073 for support.

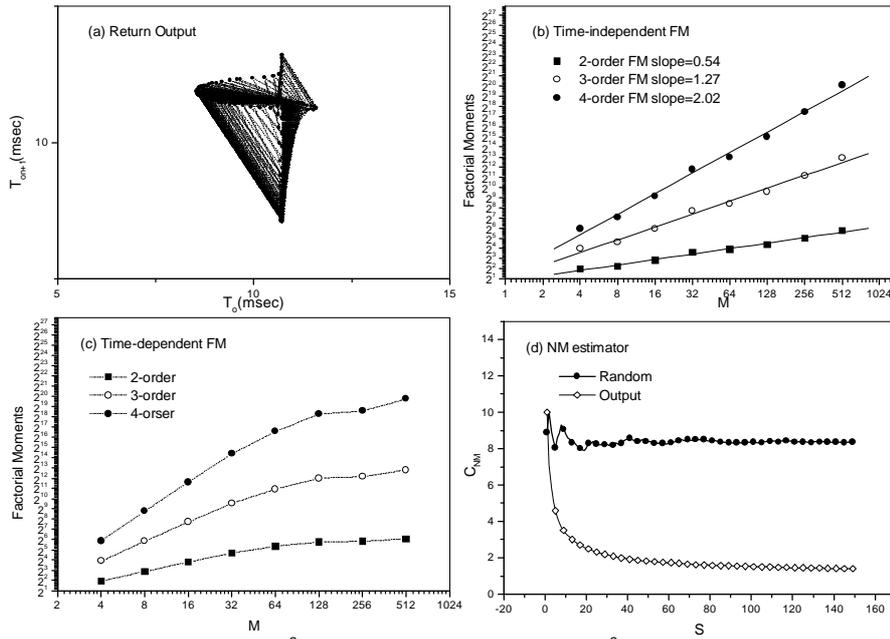

Fig.(1) Results for $I_s=25\mu A/cm^2$, $T_i=30msec$ and $g_s=0.5mS/cm^2$. (a) Return maps for output $T_o$;
(b) Time-independent factorial moments; (c) Time-dependent factorial moments; (d) NM estimator

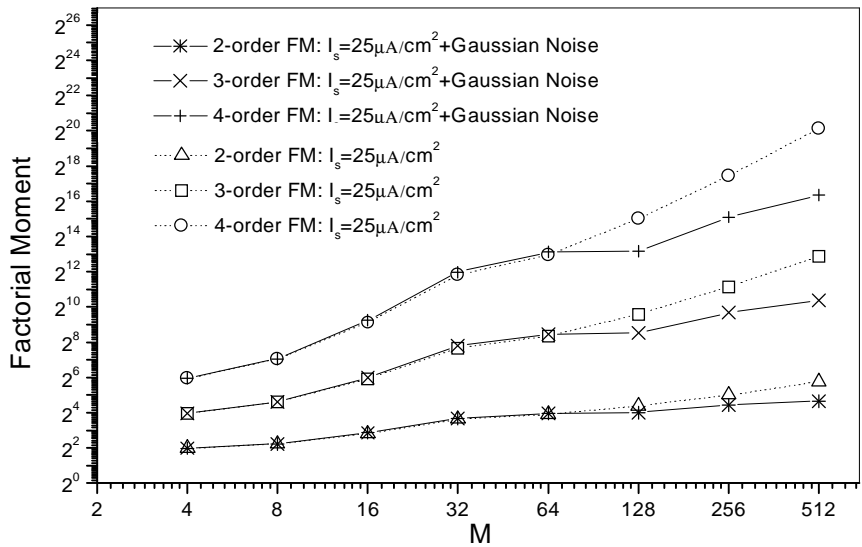

Fig.(2) Effectiveness of external noises on factorial moments. $T_i=30msec$